\documentclass[12pt,a4paper]{article}
\usepackage[left=2.5cm,right=2.5cm,top=2.5cm,bottom=2.5cm]{geometry}
\usepackage{amssymb}
\usepackage{amsmath}
\usepackage{epsfig}

\renewcommand\l[1]{\label{eq:#1}}
\renewcommand\r[1]{(\ref{eq:#1})}
\newcommand\x{\times}
\renewcommand\.{\cdot}
\renewcommand\_[1]{{\bf #1}}
\renewcommand\=[1]{\overline{\overline{#1}}}
\renewcommand\^[1]{\widehat{#1}}
\newcommand\vect[1]{\left\lgroup\!\!\begin{array}{r}#1\end{array}\!\!\right\rgroup}
\newcommand\matr[1]{\left\lgroup\!\!\begin{array}{cc}#1\end{array}\!\!\right\rgroup}

\title{Symmetry and reciprocity constraints on diffraction by
  gratings of quasi-planar particles}

\author{S.I.  Maslovski$^1$, D.K.\ Morits$^1$, S.A.\ Tretyakov$^2$\\
$^1$Radiophysics Dept.,
St.~Petersburg State Polytechnical Univ.\\
Polytechnicheskaya 29,
195251, St.~Petersburg, Russia\\ Email: {stanislav.maslovski@gmail.com}\\
$^2$Dept.\ of Radio Science and Engineering,
Helsinki Univ.\ of Technology\\
P.O.\ Box 3000,
FI-02015 TKK, Finland\\ Email: {sergei.tretyakov@tkk.fi} 
}

\begin{document}

\maketitle

\begin{abstract}
Symmetry and reciprocity constraints on polarization state of the field
diffracted by gratings of quasi-planar particles are considered.
It is shown that the optical activity effects observed
recently in arrays of quasi-planar plasmonic particles on a dielectric
substrate are due to the reflection of the field at the air-dielectric
slab interface and are proportional to this reflection coefficient.
\end{abstract}

\section{Introduction}

Two-dimensional periodical grids of thin planar metal particles of various complex
shapes show interesting electromagnetic properties and find applications
in radio and microwave engineering as frequency selective surfaces and polarization transformers (e.g., \cite{Munk}).
With recent advances in nanofabrication, optical properties of such
structures start attracting considerable interest of researchers. In particular interest are
2D-chiral inclusions, that do not possess mirror symmetry under
two-dimensional reflections with respect to a line in the particle plane. Various
gammadion shapes have been recently studied theoretically and
experimentally in the optical region, e.g.,  \cite{Zheludev1,Zheludev2,giant}.

The symmetry of these shapes has an interesting analogy with
magnetized ferrites and with three-dimensional (volumetric) chiral
particles. The gammadion (swastika) geometry defines an axial unit
vector normal to the particle plane, whose direction is determined
by the tilt direction of the particle arms and the right-hand rule.
Similarly, in the case of a magnetized ferrite sample, there is a
selected direction defined by the bias magnetic field. At first
sight, this suggests that in arrays of 2D-chiral particles there can
be electromagnetic phenomena analogous to the Faraday rotation of
the polarization plane. And indeed, polarization rotation (optical
activity) has been observed in arrays of swastika-shaped particles
on a dielectric \cite{Zheludev1,Zheludev2,giant}. However,  in
\cite{Zheludev2} this phenomenon was mistakenly attributed to broken
time-reversal symmetry or nonreciprocity (note that the Faraday
effect is nonreciprocal). Later it was experimentally shown that the
rotation of the polarization plane of transmitted waves was
reciprocal \cite{giant}. Moreover, in \cite{giant} the dielectric
substrate was found to be the key factor for the optical activity in
quasi-planar arrays of gammadions because adding a substrate would
effectively turn an array of 2D-chiral objects into a 3D-chiral
structure which would exhibit the usual reciprocal optical activity.

Despite apparent similarity, polarization rotation phenomena are fundamentally different in reciprocal and nonreciprocal
systems. In the case of magnetized ferrites, time reversal operation reverses the direction of the
axial vector defining the bias field direction, while in the case of
arrays of reciprocal particles, time reversal does not affect the direction of the
axial vector showing the particle ``rotation sense''. Any system formed by
metal and dielectric inclusions of any shapes is reciprocal, because there is no
time-odd physical parameter there and the Maxwell equations themselves are invariant
under time reversal operation.

It is well known that polarization plane rotation
exists also in reciprocal media. Optical activity is the main
manifestation of three-dimensional chirality of molecules or
artificial inclusions forming chiral and, more generally,
bi-anisotropic materials \cite{chibi,biama}. This effect is governed
by the chirality parameter (the trace of the magnetoelectric
coupling dyadic), which vanishes for planar particles of arbitrary
shapes, because any 2D shape can always be superimposed with its
mirror image if rotations in 3D space are allowed. In fact, the magnetoelectric coupling
dyadic of planar particles is antisymmetric, and composites formed by such particles belong to the
class of omega media \cite{biama}.  This
apparently imposes restrictions on optical activity effects in
arrays of 2D particles, even if their shape is chiral in the
two-dimensional space.

Electromagnetic properties of 2D-chiral arrays in the microwave region were
studied earlier (e.g. \cite{Luk,old1}) and these structures were even patented
\cite{Said_patent}. However, to the best of our knowledge,
no systematic consideration of fundamental limitations on electromagnetic
scattering from general grids of quasi-planar particles has been made.
Important results were obtained in \cite{Luk,Dmitriev_EL,Dmitriev2} from the symmetry considerations,
but those results concern restrictions on the form of the constitutive tensors of
composite media but not the diffraction phenomena. The theory of
 \cite{Prosvirnin}
concerns only arrays in free space and does not include the effects of the
substrate.

In this paper we make a systematic study of reciprocity and symmetry
restrictions on plane-wave diffraction by 2D arrays of arbitrary
shaped quasi-planar particles. By quasi-planar particles we
understand particles of a finite thickness but with uniform properties
over the full particle thickness. The arrays can be positioned on
planar dielectric substrates, and we will see that the presence of the
substrate is of key importance for optical activity phenomena in these structures.

\section{Problem formulation}

\begin{figure}[h!]
\centering
\epsfig{file=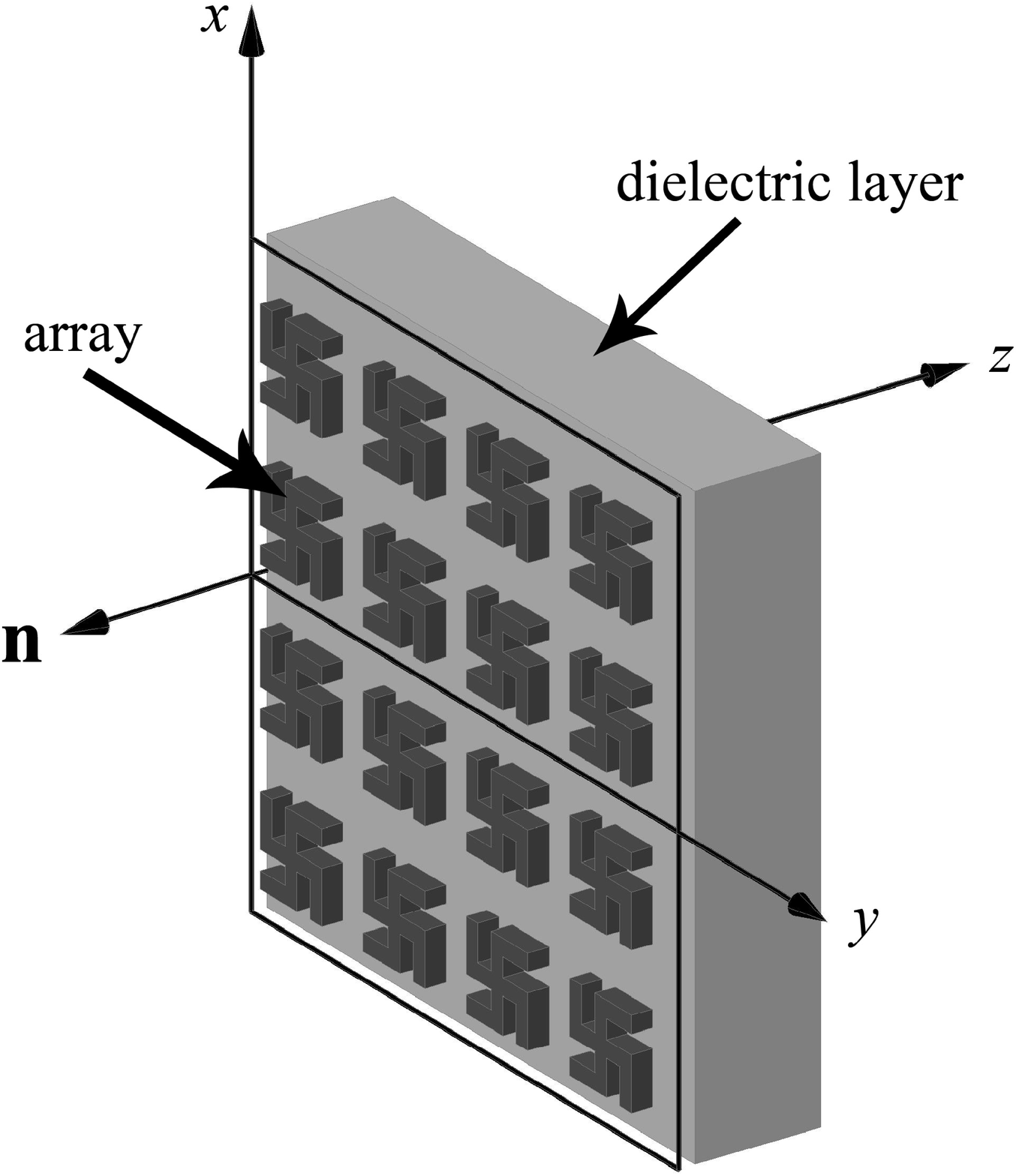,width=0.7\textwidth}
\caption{An array of metal (plasmonic) particles backed by a dielectric
  slab. $\_n$ is the unit vector normal to the array plane.}
\label{fig1}
\end{figure}

We consider an infinite two-dimensional array of arbitrary
shaped (in the array plane) particles backed with a slab of a homogeneous dielectric. The
thickness of the particles is assumed to be finite, but the structure is
uniform over its thickness. This implies that the array has a plane of
symmetry that itself can be regarded as the array plane. In practice, the particles are usually made of metal,
but the theory which we develop here applies to any reciprocal isotropic material.
In case of metal inclusions, the general theory developed here applies both to microwave frequencies and to the visible
range, because particular electromagnetic parameters of metal particles
do not affect restrictions which follow from symmetry and reciprocity considerations.
We assume that the array particles are not touching the
dielectric substrate, while the separation of the dielectric and the particles
can be arbitrary small. This assumption allows us to split the entire
structure in two parts: the grating and the slab and consider them separately.
An example of the problem geometry is illustrated in Fig.~\ref{fig1}.

The array is illuminated by a plane wave of arbitrary polarization and arbitrary propagation
direction\footnote{Evanescent-wave excitation is not excluded from this analysis.} incoming from the half space
$z < 0$. We are interested in the polarization states of the
transmitted and reflected waves\footnote{The case of multiple propagating modes created by the
array is not excluded: In that case the following restrictions apply to every
excited mode.} and how it relates to the polarization
of the incident wave.
We will attack this problem with a rather general approach taking
into account symmetry and reciprocity restrictions.

\section{Symmetry and reciprocity restrictions}

\begin{figure}[h!]
\centering
\epsfig{file=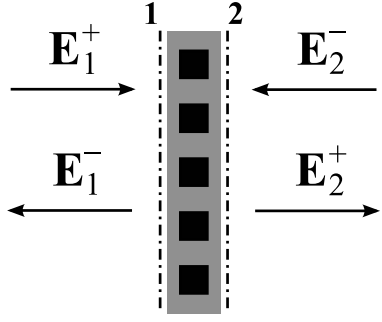,width=0.3\textwidth}
\caption{Definitions of the incident and scattered waves for a quasi-planar grid. The arrows show the
normal to the interfaces components of the wave vector.}
\label{Interfaces}
\end{figure}

Let us start from considering the array of particles separately from the dielectric substrate.
Under our assumptions there exist two planes in space which enclose the grating entirely in
between themselves. We split the tangential electric fields on
these two planes into the incident fields exciting the grating and the
fields scattered by the grating (see Fig.~\ref{Interfaces}) and introduce
the following scattering matrix:
\begin{equation}\vect{
\_E_1^-(\_r)\\
\_E_2^+(\_r)} =
\matr{
\^S_{11}& \^S_{12}\\
\^S_{21}& \^S_{22}} \.
\vect{
\_E_1^+(\_r)\\
\_E_2^-(\_r) }.
\l{opS} \end{equation} Here $\_E_1^+(\_r)$ and $\_E_2^-(\_r)$
denote the vectorial complex amplitudes of the tangential components
of the incident electric field on the first and the second
surface, respectively. $\_E_1^-(\_r)$ and $\_E_2^+(\_r)$ are the
respective scattered fields, and $\^S_{mn}$ are scattering operators
that connect the tangential components of the scattered and incident
electric fields at surfaces 1 and 2.

The fields at the two surfaces can be expanded into plane waves
\begin{equation}\_E(\_k_{\rm t}) =
\iint \_E(\_r)\,e^{-i\_{\_k_{\rm t}}\.\_r}d^2\_r,
\end{equation} where the integration is taken over the surface plane, $\_E$ can be
any of the fields defined above, and $\_k_{\rm t}$ is the wave vector component tangential to
the planes. We use the time dependence of the form $\exp(-i\omega t)$ conventionally used in optics.

This expansion gives us a representation of the operators
$\^S_{mn}$ in the basis of plane waves. The corresponding matrix
elements of the operators $\^S_{mn}$ in this basis are dyadics $\=s_{mn}(\_p,\_q)$:
\begin{equation}\=s_{mn}(\_p,\_q) = \iint
e^{-i\_p\.\_r}\,\^S_{mn}\,e^{i\_q\.\_r}d^2\_r.
\l{matr_el}
\end{equation}
If the grating has a center of symmetry and the planes where we
define the fields are located at symmetric positions with respect to the
same center, the dyadics $\=s_{mn}(\_p,\_q)$ obey
the following restrictions:
\begin{equation}\=s_{11}(\_p,\_q) = \=s_{22}(-\_p,-\_q),\quad \=s_{12}(\_p,\_q) =
\=s_{21}(-\_p,-\_q).
\l{centsym}
\end{equation}
Indeed, applying the coordinate transform $\_r \rightarrow -\_r$ where
the origin is at the symmetry
center, we come to the same structure with the first and the
second interfaces interchanged. On the other hand, the components of polar
vectors change signs under this transformation, and we obtain~\r{centsym}.

If the grating plane is itself a plane of symmetry of the system,
then, more obviously,
\begin{equation}\^S_{11}=\^S_{22}, \quad \^S_{12} = \^S_{21},
\l{planesym}
\end{equation} or, which is the same,
\begin{equation}\=s_{11}(\_p,\_q) = \=s_{22}(\_p,\_q),\quad \=s_{12}(\_p,\_q) =
\=s_{21}(\_p,\_q).
\l{planesym_el}
\end{equation} To get this result, one has to apply a transform that changes the sign of the
radius vector component normal to the grating plane. Such a transformation interchanges
the interfaces as above, but preserves the signs of the tangential components
of polar vectors.

Furthermore, if the grating has rotational symmetry of order $m \ge 2$ with respect to
an axis parallel to $\_n$, the following additional relation holds:
\begin{equation}\=s_{ij}(\=\xi\.\_p,\=\xi\.\_q) =
\=\xi\.\=s_{ij}(\_p,\_q)\.{\=\xi\,}^{-1}.
\l{rot_sym}
\end{equation} where $\=\xi = \exp[({2\pi/m)[\_n\x\=I_{\rm t}]}]$ is the dyadic of
rotation in the grating plane, $\=I_{\rm t}$ is the unit dyadic in the same plane.

In a reciprocal system the reciprocity theorem holds. To apply this theorem we assume that
the external fields are created by sheets of surface electric
currents $\_J_{1,2}$ placed at the first and the second interfaces,
respectively. Then, at the interfaces
\begin{equation}\_E_1^+(\_r) = -{\^Z_{\rm w}\.\_J_1(\_r)\over 2}, \quad \_E_2^-(\_r) = -{\^Z_{\rm w}\.\_J_2(\_r)\over 2},
\l{ext_field}
\end{equation} where $\^Z_{\rm w}$ is the free-space wave impedance operator. In the
plane-wave basis it is
\begin{equation}\=z_{\rm w}(\_p,\_q) =
(2\pi)^2\,\delta(\_q-\_p)\,\=z_{\rm w}(\_q), \quad
\=z_{\rm w}(\_q) = \eta\,{\=I_{\rm
    t}-\_q\_q/k^2\over\sqrt{1-q^2/k^2}},
\l{zw_el}
\end{equation} where $\eta = \sqrt{\mu_0/\varepsilon_0}$, $k =
\omega\sqrt{\varepsilon_0\mu_0}$. The operator $\^Z_{\rm w}$ is
diagonal and symmetric as is directly seen from \r{zw_el}:
\begin{equation}\=z_{\rm w}(\_p,\_q) = \=z_{\rm w}^{\rm T}(-\_q,-\_p); \quad
\=z_{\rm w}(\_p,\_q) = 0, \ \mbox{if } \_p\neq\_q.
\l{zw_diag}
\end{equation}
Because the external field itself satisfies the reciprocity theorem, we can
write for the scattered fields
\begin{equation}\iint\_E_{1'}^-\.\_J_1\,d^2\_r + \iint\_E_{2'}^+\.\_J_2\,d^2\_r =
\iint\_E_{1}^-\.\_J_{1'}\,d^2\_r + \iint\_E_{2}^+\.\_J_{2'}\,d^2\_r.
\l{recip_th}
\end{equation} where $\_J_{1,1'}$, $\_J_{2,2'}$ are the external currents and
$\_E_{1,1'}^-$, $\_E_{2,2'}^+$ are the scattered fields in a pair of
separate excitation scenaria.

The currents $\_J_{1,2}$, $\_J_{1',2'}$ in \r{recip_th} can be
arbitrary chosen and the corresponding electric fields can be
expressed from \r{opS} and \r{ext_field}. If, for example, $\_J_2 = \_J_{2'} = 0$, we get
from~\r{opS} and~\r{ext_field}
\begin{equation}\iint \_J_1\.(\^S_{11}\.\^Z_{\rm w})\.\_J_{1'}\,d^2\_r =
\iint \_J_{1'}\.(\^S_{11}\.\^Z_{\rm w})\.\_J_1\,d^2\_r,
\end{equation} which in the operator sense means that transpose does not change the
operator in braces. Therefore, considering all possible cases, we
get for a reciprocal grating:
\begin{equation}\^S_{11}\.\^Z_{\rm w}=\^Z_{\rm w}\.\^S_{11}^{\rm T}, \quad
\^S_{22}\.\^Z_{\rm w}=\^Z_{\rm w}\.\^S_{22}^{\rm T}, \quad
\^S_{12}\.\^Z_{\rm w}=\^Z_{\rm w}\.\^S_{21}^{\rm T}.
\l{recipr_a}
\end{equation} Here we use the fact that $\^Z_{\rm w} = \^Z_{\rm w}^{\rm T}$ as
given by \r{zw_diag}. Relations
\r{recipr_a} can be also understood as transpose rules
\begin{equation}\^S_{11}^{\rm T}=\^Z_{\rm w}^{-1}\.\^S_{11}\.\^Z_{\rm w}, \quad
\^S_{22}^{\rm T}=\^Z_{\rm w}^{-1}\.\^S_{22}\.\^Z_{\rm w}, \quad
\^S_{21}^{\rm T}=\^Z_{\rm w}^{-1}\.\^S_{12}\.\^Z_{\rm w}.
\l{recipr}
\end{equation}
Eq.~\r{recipr} can be rewritten in terms of the matrix elements of the
corresponding operators:
\begin{equation}\begin{split}
\=s_{11}^{\rm T}(-\_q,-\_p) &=
\=z_{\rm w}^{\,-1}(\_p)\.\=s_{11}(\_p,\_q)\.\=z_{\rm w}(\_q),\\
\=s_{22}^{\rm T}(-\_q,-\_p) &=
\=z_{\rm w}^{\,-1}(\_p)\.\=s_{22}(\_p,\_q)\.\=z_{\rm w}(\_q),\\
\=s_{21}^{\rm T}(-\_q,-\_p) &=
\=z_{\rm w}^{\,-1}(\_p)\.\=s_{12}(\_p,\_q)\.\=z_{\rm w}(\_q).
\end{split}
\l{recipr_el}
\end{equation}
From \r{recipr_el} one can see that for the waves propagating normally
to the grating (when $\_p=\_q=0$) reciprocity simply requires
\begin{equation}\=s_{11}(0,0) = \=s_{11}^{\rm T}(0,0), \quad
\=s_{22}(0,0) = \=s_{22}^{\rm T}(0,0), \quad
\=s_{12}(0,0) = \=s_{21}^{\rm T}(0,0).
\l{recipr_norm}
\end{equation}
Let us note that when compared to the standard mode coupling
parameters in the waveguide theory (S-parameters), our scattering
dyadics $\=s_{mn}(\_p,\_q)$ differ in that sense that, first, we do
not decompose the fields into TE and TM modes, and, second, we do not
separately normalize the modes, instead we deal with a simple plane-wave
basis applicable to all types of modes. This allows for a dyadic
formalism but results in additional impedance terms in~\r{recipr_el}.

\section{Multiple layers}

When there are several layers stacked one on top of another (in
our case there are two layers: the grating and the dielectric) the
direct application of $\^S$-matrix \r{opS} becomes tedious. Instead, it
is better to apply an approach based on transfer matrices. A
transfer matrix connects pairs of fields given at separate
interfaces:
\begin{equation}\vect{
\_E_1^+(\_r)\\
\_E_1^-(\_r)
} =
\matr{
\^T_{11} & \^T_{12}\\
\^T_{21} & \^T_{22}
}\.
\vect{
\_E_2^+(\_r)\\
\_E_2^-(\_r)
}.
\end{equation} The total transfer matrix of a layered structure is a product of
transfer matrices of the layers. The operators $\^T_{mn}$ can be
expressed trough $\^S_{mn}$ and {\it vice versa:}
\begin{equation}\^T_{11} = \^S_{21}^{-1}, \quad
\^T_{12} = -\^S_{21}^{-1}\.\^S_{22}, \quad
\^T_{21} = \^S_{11}\.\^S_{21}^{-1}, \quad
\^T_{22} = \^S_{12} - \^S_{11}\.\^S_{21}^{-1}\.\^S_{22},
\end{equation} \begin{equation}\^S_{11} = \^T_{21}\.\^T_{11}^{-1}, \quad
\^S_{12} = \^T_{22} - \^T_{21}\.\^T_{11}^{-1}\.\^T_{12}, \quad
\^S_{21} = \^T_{11}^{-1}, \quad
\^S_{22} = -\^T_{11}^{-1}\.\^T_{12}.
\end{equation}
From these rather complicated relations one can see that the
symmetries considered in the previous section generally do not result in
simple constraints on $\^T_{mn}$ (when compared to
$\^S_{mn}$). This explains why we have started our formulation
from the matrix of $\^S_{mn}$ operators. One exception is
the rotational symmetry rule \r{rot_sym} which also holds
for~$\^T_{mn}$.

However, there is one important case which results in simple
constraints on the elements of the transfer matrix. It is the case when
the structure is reciprocal and has a plane of symmetry, i.e. its
$\^S_{mn}$ operators satisfy both \r{planesym} and \r{recipr}.
In this case after some operator algebra we obtain
\begin{equation}\^T_{11}^{\rm T} = \^Z_{\rm w}^{-1}\.\^T_{11}\.\^Z_{\rm w}, \quad
\^T_{22}^{\rm T} = \^Z_{\rm w}^{-1}\.\^T_{22}\.\^Z_{\rm w}, \quad
\^T_{12}^{\rm T} = -\^Z_{\rm w}^{-1}\.\^T_{21}\.\^Z_{\rm w}.
\l{symm_T}
\end{equation} In the opposite direction, if the operators $\^T_{mn}$ satisfy
\r{symm_T} then the operators $\^S_{mn}$ are such that
\begin{equation}\^S_{11}^{\rm T} = \^Z_{\rm w}^{-1}\.\^S_{22}\.\^Z_{\rm w}, \quad
\^S_{12}^{\rm T} = \^Z_{\rm w}^{-1}\.\^S_{12}\.\^Z_{\rm w}, \quad
\^S_{21}^{\rm T} = \^Z_{\rm w}^{-1}\.\^S_{21}\.\^Z_{\rm w}.
\l{symm_S}
\end{equation} Eqs.~\r{symm_T}, \r{symm_S} can be rewritten in terms of the matrix
elements of the corresponding operators analogously to \r{recipr_el}.

\section{Applications}

In the previous sections we formulated a number of symmetry and
reciprocity constraints on scattering and transfer matrices.
Let us consider some consequences of these constraints.

\subsection{Restrictions on optical activity}

First, it is easy to see that the combination of the reciprocity and
the central symmetry in a structure forbids any optical activity
effects for normally incident plane waves (both in transmission and
reflection into the same normally propagating mode).
Indeed, from~\r{centsym} and~\r{recipr_norm} we get
\begin{equation}\begin{split}
\=s_{11}(0,0) = \=s_{22}(0,0) =
\=s_{11}^{\rm T}(0,0) = \=s_{22}^{\rm T}(0,0),\\
\=s_{21}(0,0) = \=s_{12}(0,0) =
\=s_{21}^{\rm T}(0,0) = \=s_{12}^{\rm T}(0,0),
\end{split}
\end{equation} i.e. both transmission and reflection dyadics are fully symmetric. For
optical activity to occur one needs antisymmetric terms. We have
the same result if the central symmetry is replaced by the planar
symmetry~\r{planesym_el}.
Moreover, if the structure also has rotational symmetry~\r{rot_sym}
with $m \ge 3$ then $\=s_{mn}(0,0) = s_{mn}\=I_{\rm t}$, i.e. the structure does not
change the polarization state of the transmitted and reflected fields
at all.

Next, let us look at the original problem of an array of
quasi-planar 2D-chiral
particles on top of a dielectric slab. To obtain the total transfer
matrix of this system one has to multiply the transfer matrix of the
grating by the transfer matrix of the slab:
\begin{equation}\matr{\^T_{mn}^{\rm\,tot}} =
\matr{\^T_{11} & \^T_{12}\\
\^T_{21} & \^T_{22}} \.
\matr{\^{\cal T}^{-1} & -\^{\cal T}^{-1}\.\^{\cal R}\\
\^{\cal R}\.\^{\cal T}^{-1} & (\^{\cal T} - \^{\cal R}\.\^{\cal
  T}^{-1}\.\^{\cal R})},
\l{tot_tm}
\end{equation} where $\^{\cal T}$ and $\^{\cal R}$ denote the transmission and
reflection operators of the slab, respectively. The matrix elements of
these operators are dyadics $\=\tau(\_p,\_q)$ and $\=\rho(\_p,\_q)$:
\begin{equation}\=\tau(\_p,\_q) = (2\pi)^2\,\delta(\_q-\_p)\,\=\tau(\_q), \quad
\=\rho(\_p,\_q) = (2\pi)^2\,\delta(\_q-\_p)\,\=\rho(\_q),
\l{tau_rho}
\end{equation} where $\=\tau(\_q)$ and $\=\rho(\_q)$ are dyadic transmission and
reflection coefficients for a dielectric slab as functions of the
transversal wave vector of the incident wave ($\_k_t = \_q$). The explicit expressions for
them can be found in textbooks on electromagnetics.

We are interested in $\^T_{11}^{\rm\,tot}$ because it is connected with
the transmission operator of the total system: $\^S_{21}^{\rm\,tot} =
(\^T_{11}^{\rm\,tot})^{-1}$. From \r{tot_tm} we get
\begin{equation}\^T_{11}^{\rm\,tot} =
\^T_{11}\.\^{\cal T}^{-1} + \^T_{12}\.\^{\cal R}\.\^{\cal T}^{-1},
\end{equation} or, rewriting it in terms of the matrix elements,
\begin{equation}\=t^{\rm\,tot}_{11}(\_p,\_q) = \=t_{11}(\_p,\_q)\.\=\tau^{\,-1}(\_q) +
\=t_{12}(\_p,\_q)\.\=\rho(\_q)\.\=\tau^{\,-1}(\_q).
\l{t11_tot}
\end{equation}
Let us now look at the transpose of the operator
$\^T_{11}^{\rm\,tot}$ in the case when the grating is
reciprocal and has a plane of symmetry. In this case
\begin{equation}(\^T_{11}^{\rm\,tot})^{\rm T} =
\^{\cal T}^{-1}\.\^Z_{\rm w}^{-1}\.\^T_{11}\.\^Z_{\rm w} -
\^{\cal T}^{-1}\.\^{\cal R}\.\^Z_{\rm w}^{-1}\.\^T_{21}\.\^Z_{\rm
  w},
\l{T11_tot_t}
\end{equation} where we used \r{symm_T} and the fact that the transmission and
reflection operators of a dielectric slab are diagonal and symmetric:
$\^{\cal T} = \^{\cal T}^{\rm T}$,
$\^{\cal R} = \^{\cal R}^{\rm T}$.
In terms of the matrix elements \r{T11_tot_t} becomes
\begin{equation}{\=t_{11}^{\rm\,tot}}^{\rm T}(-\_q,-\_p) =\\
\=\tau^{\,-1}(\_p)\.\left[
\=z_{\rm w}^{\,-1}(\_p)\.\=t_{11}(\_p,\_q) -
\=\rho(\_p)\.\=z_{\rm w}^{\,-1}(\_p)\.\=t_{21}(\_p,\_q)\right]\.\=z_{\rm w}(\_q).
\l{t11_tot_t}
\end{equation} Eqs.~\r{t11_tot} and \r{t11_tot_t} greatly simplify for the plane waves propagating
along the normal:
\begin{eqnarray}
\=t_{11}^{\rm\,tot}(0,0) &=& \tau^{-1}\,\=t_{11}(0,0) + \rho\,\tau^{-1}\,\=t_{12}(0,0),\\
{\=t_{11}^{\rm\,tot}}^{\rm T}(0,0) &=&
\tau^{-1}\,\=t_{11}(0,0) - \rho\,\tau^{-1}\,\=t_{21}(0,0),
\end{eqnarray}
where $\rho$ and $\tau$ are scalar reflection and transmission
coefficients for a plane wave normally incident on a dielectric slab.
From the last relations one can obtain an explicit formula for the
antisymmetric part of $\=t_{11}^{\rm\,tot}(0,0)$:
\begin{equation}{{\=t_{11}^{\rm\,tot}}(0,0) - {\=t_{11}^{\rm\,tot}}^{\rm T}(0,0)\over 2} =
\rho\,{\=t_{12}(0,0) + \=t_{21}(0,0)\over 2\,\tau} =
\rho\,{\=t_{12}(0,0) - \=t_{12}^{\rm T}(0,0)\over 2\,\tau}.
\l{antisym_t}
\end{equation}
From here one can see that adding a dielectric slab to a reciprocal
symmetric grating can indeed result in a system with
a non-symmetric transmission dyadic, i.e. one can realize an optically
active system this way. The higher is the reflection, the higher is
the obtained optical activity. The physical reason for this is that
the dielectric layer positioned on one side of the array breaks the
mirror symmetry of the array, and the whole structure becomes chiral
in the three-dimensional space.

Furthermore, we see that the optical activity of this kind cannot be achieved
with gratings that have $\=t_{12}(0,0) + \=t_{21}(0,0) = 0$. For
example, this holds for any planar (thickness tends to zero)
grating made of a non-magnetic material. Indeed, for such a grating
\begin{equation}\^S_{22} = \^S_{11}, \quad \^S_{21} = \^S_{12} = 1 + \^S_{11},
\end{equation} therefore
\begin{equation}\^T_{12} = -(1 + \^S_{11})^{-1}\.\^S_{11} = -(1 + \^S_{11}^{-1})^{-1}
= -\^S_{11}\.(1 + \^S_{11})^{-1} = -\^T_{21}.
\end{equation}
This observation leads us to an additional conclusion that the
obtained optical activity depends also on the thickness of the
quasi-planar grating. Increasing the thickness (up to a certain limit
related to the wavelength) one can increase the
term $\=t_{12}(0,0) + \=t_{21}(0,0)$ in~\r{antisym_t}.

\subsection{An array of swastika-shaped particles}

As an illustrative example let us consider a square array of
swastika-shaped metal particles. It was found experimentally that such
a grating when backed with a dielectric slab rotates the polarization
of normally incident plane waves \cite{giant}. When there is no substrate, there will
be no rotation. Although this follows from our general considerations
given above, let us show the same using simpler arguments.

\begin{figure}[h!]
\centering
\epsfig{file=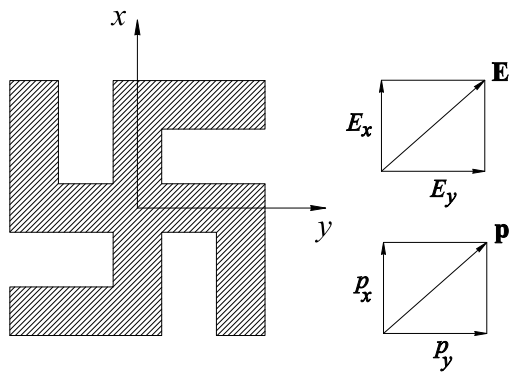,width=0.4\textwidth}
\caption{A swastika-shaped particle excited by an external electric
  field $\_E$. Here $\_p$ is the induced electric dipole moment.}
\label{fig2}
\end{figure}

Consider a swastika-shaped metallic particle depicted in Fig.~\ref{fig2}.
The particle has rotational symmetry or order 4, it has a center of
symmetry, and it is reciprocal. Suppose that the particle is under
normal incidence of
a linearly polarized plane wave with the electric field vector $\_E$
oriented as shown in the figure. One can decompose the electric field
vector of the incident wave to two components along $x$ and $y$ axes.
Due to rotational symmetry the particle reacts equally to both these
components.

The induced electric dipole moment of the particle is a sum of two
components $\_p_x$ and $\_p_y$, and the total dipole moment
$\_p = \_p_x + \_p_y$ of the particle can be written as
\begin{equation}\_p = \=a\.\_E,
\end{equation} where $\=a$ is the dyadic dipole polarizability of the
particle. The rotational symmetry requires that $a_{xx} = a_{yy}$,
$a_{xy} = -a_{yx}$. On the other hand, the reciprocity requires
$a_{xy} = a_{yx}$, therefore $a_{xy} = a_{yx} = 0$ and $\=a = a\=I_t$,
where $\=I_t$ is the two-dimensional unit dyadic. This means that
the induced dipole moment is co-linear with the applied electric
field.

In a square array the interaction of the particles will exhibit the
same rotational symmetry, therefore for the dipole moment
per unit cell of the array and the polarizability in the array
we will arrive to the same conclusions as above.
The secondary field created by the array in far zone is proportional
to the dipole moment per unit cell of the array. Its polarization
is the same as the polarization of the induced dipole moment, and,
hence, it is the same as the polarization of the incident field.

We have to note that this simplistic consideration does not
include higher multipole moments, and, what is more important, it
assumes that the effective induced magnetic moment per unit cell of
the array is either zero (as in swastikas under normal incidence)
or does not radiate in the far zone (which is the case, for instance, of
omega particles under normal incidence, when the induced magnetic
dipole moment is parallel to the propagation direction).
It is known that one has to provide some means for magnetoelectric
interactions to obtain optical activity. Adding a dielectric substrate
breaks mirror symmetry of a quasi-planar array and makes
optical activity possible, as we have shown in the previous section. A detailed
study of how this process can be described in terms of magnetoelectric
interactions is outside of the scope of this paper.

\section{Conclusions}

Starting from general symmetry and reciprocity (time-reversal invariance) requirements we have
formulated a number of constraints on the scattering and transmission
operators of arrays of quasi-planar linear reciprocal particles. We
have shown that if a reciprocal array is symmetric with respect to its own
plane or has a center of symmetry, then this array cannot exhibit
optical activity for normally incident plane waves both in
transmission or reflection.
For the case of a reciprocal plane-symmetric array backed by a dielectric slab we have
shown that the optical activity effects for normally incident waves occur due to reflection at the
air-dielectric interface and the strength of this effect is proportional to
the respective reflection coefficient.
Moreover,
even in the presence of a dielectric substrate, to allow for optical
activity the thickness of the particles in the array should not be
negligibly small.

The general restrictions on
scattering and transmission operators obtained under this research apply also to grids
that create several propagating modes in reflection and transmission
(electrically sparse arrays producing diffraction lobes).
In the last case, the restrictive relations become more complicated and
they allow for asymmetries of diffraction phenomena under reversal of propagation directions.
However, all these phenomena are reciprocal and agree with the
fundamental time-reversal invariance of the Maxwell equations.

\end{document}